\documentclass{ieeeaccess}
\usepackage{cite}
\usepackage{amsmath,amssymb,amsfonts}
\usepackage{algorithmic}
\usepackage{graphicx}
\usepackage{textcomp}
\usepackage{color}
\usepackage{soul}

\def\hl#1{#1} 

\usepackage{refcount} 
\usepackage{cases}
\usepackage{alltt}
\usepackage{hyperref}

\allowdisplaybreaks

\begin{document}
\history{Fourth version,  22 October, 2021}
\doi{DOI to be assigned}
\title{Comment on ``Bell’s Theorem Versus Local Realism in a Quaternionic Model of Physical Space''}
\author{\uppercase{Richard D.~Gill}\authorrefmark{1}}
\address[1]{Mathematical Institute, Leiden University, Netherlands (e-mail: gill@math.leidenuniv.nl)}
\tfootnote{No financial support was obtained for this research}

\markboth
{Richard D. Gill: Comment on ``Bell’s Theorem Versus Local Realism...''}
{Richard D. Gill: Comment on ``Bell’s Theorem Versus Local Realism...''}

\date{22 October 2021; fourth version}                                           

\begin{abstract}
\noindent I point out critical errors in the paper ``Bell’s Theorem Versus Local Realism in a Quaternionic Model of Physical Space'' by J. Christian, published in {\em  IEEE Access}. Christian's paper in fact contains several conflicting models. None of them form counterexamples to Bell's theorem. Most of Christian's paper is devoted to a model based on the detection loophole due to Pearle (1970).
\end{abstract}


\titlepgskip=-15pt

\maketitle

\section{Introduction}

\noindent  Bell's 1964 theorem \cite{Bell} states that the conventional framework of quantum mechanics is incompatible with a physical principle called \emph{local realism}. Bell's theorem is a cornerstone of modern quantum information theory, and of quantum computing. Proof that it is wrong would unleash a revolution in science with enormous impact on society and technology. Every textbook on quantum mechanics would have to be rewritten.

At the core of Bell's proof of his theorem is an elegant and simple probability inequality, going back to Boole (1853) \cite{Boole}. However, Christian (2019) \cite{IEEEAccess1} claims in \emph{IEEE Access} that the usual proof of this inequality depends on an incorrect physical assumption. He goes on to present two quite different models which he claims form counterexamples to the inequality and hence to Bell's theorem. His main tool is Geometric Algebra, see Doran and Lasenby (2003) \cite{DoranLasenby}.

Christian's 2019 paper was not his first publication with this theme, nor the last. The present author surveyed Christian's work on Bell's theorem from 2007 to 2019 in \cite{gill2020a}, thus already including a section on the paper which is the subject of the present ``Comment''. Soon after that, I published a ``Comment'' \cite{gill2021} to Christian's (2020) companion \emph{IEEE  Access} paper \cite{IEEEAccess2}. I showed in \cite{gill2020a} and \cite{gill2021} that Christian's results are based on elementary errors in algebra, calculus and reasoning. Bell's theorem has not been disproved and Christian's thesis that the quantum correlations have their origin in the geometry of space is not supported. That is not to say that his claim might not be true. Christian makes an intriguing suggestion, but that is all.

{Christian has responded with a ``Reply'' to my ``Comment''} \cite{gill2021} {on his companion paper. Obviously, readers of \emph{IEEE  Access} must judge for themselves whether or not he has refuted my criticisms. Here, I would just like to comment on some revealing remarks from} \cite{christian2021}. {Christian writes}
\begin{quote}
{That is not to say that Bell's theorem does not have a sound mathematical core. 
When stated as a mathematical theorem in probability theory, there can be no doubt about its validity. 
But my work on the subject does not challenge this mathematical core, if it is viewed as a piece of mathematics.
What it challenges are the metaphysical conclusions regarding locality and realism derived from that mathematical core.
My work thus draws a sharp distinction between the mathematical core of Bell’s theorem and the metaphysical conclusions derived from it.}
\end{quote}
{I wish here to respond to these remarks as follows.
Firstly, Christian has claimed again and again to have (yet another) \emph{mathematical counterexample} to Bell's core \emph{mathematical theorem}. 
In other words, in one paper after another, he proposes a mathematical model which, he claims, satisfies the mathematical assumptions used to derive the CHSH inequality, but which still violates that inequality.
Secondly, he claims that his work \emph{draws a sharp distinction} between the mathematical core of Bell's theorem and metaphysical conclusions derived from it. That is a noble intention, but, in my opinion, his work conspicuously \emph{fails} to draw that distinction}.

In this paper, I will focus on the main conceptual errors in \cite{IEEEAccess1} and also explain how Christian's computer simulation manages to achieve the impossible. The simulation described in Section VI of \cite{IEEEAccess1} is simply an implementation of Pearle's (1970) \cite{pearle} detection loophole model. The idea is that two particles travel to two distant detectors where they each interact with a detector. Bell's theorem makes the assumption that the interaction always leads to an outcome $\pm 1$. In a local realistic model, this outcome is modelled as a function of an externally provided setting (typically an angle, specifying a direction or an orientation) and of hidden variables associated with source and measurement apparatus. However, in real world experiments, it used not to be possible to arrange that all particles were detected. Expressed in another way, the outcomes of measurement are elements of $\{-1, 0, +1\}$ where ``0'' stands for ``no show'', no particle is detected. Pearle showed that by rejecting any trials in which either of the two outcomes was ``no show'', one could produce a selectively chosen subsample in which the probabilities of the outcomes given the settings did exactly reproduce those predicted by quantum mechanics for the famous EPR-B model.
 
The new generation of loophole-free Bell experiments \cite{Hensen2015} measure correlations between four binary variables: two binary inputs and two binary outputs; one input and output at each of two distant locations. The basic experimental unit is not ``detected particle pair'' but ``time slot''. There is no post-selection.  Bell does not take account of the geometry of space because his argument, on the side of local realism, does not depend on it in any way whatsoever. {To be sure, it is essential to Bell's analysis that something akin to the relativistic notion of spacelike separated events in spacetime makes sense. Maybe the geometry of spacetime (not space alone) could be such that there is no meaningful distinction between local common causes and nonlocal influences. Technically, this might still amount to violating the locality or the no-conspiracy assumption (while giving a geometric explanation for doing so). But Christian's speculations about a quaternionic space seem completely irrelevant.}

The present ``Comment'' contains no new results on Bell's theorem. Most of the criticism of Christian's work which I give here has appeared in earlier work either by the present author or by others. In particular, Lasenby (2020) \cite{lasenby} shows that a central and purely algebraic result in Christian (2018) \cite{RSOS} (published in \emph{Royal Society Open Science}) is wrong. Lasenby concentrates on the Geometric Algebra component of that paper. He writes ``Christian's work has repeatedly been criticised mathematically, but he has several times stated that no one well-versed in Geometric Algebra has explicitly criticised his mathematics in print, and that this suggests his critics simply do not understand the GA in his work, not that his mathematics is wrong.'' Lasenby identifies exactly the same GA errors as I did in my papers \cite{gill2020a}, \cite{gill2021}; recall that the first of those two also contains an analysis of the paper by Christian under discussion here. 

Because of this, I will not pay attention to the Geometric Algebra in Christian's paper. I will first give a short review of the Bell-CHSH inequality, the usual route to proving Bell's theorem. Christian claims to have pinpointed an error in the argument; I will refute his claim. My review will underscore the fact that the geometry of space is simply \emph{irrelevant} to  Bell's argument. I will then discuss the computer simulation in \cite{IEEEAccess1}. Christian claims that it is an implementation of his novel physical model, and that it proves that Bell's theorem is wrong. However, it is easy to see that the simulation has almost nothing to do with Christian's model. Moreover, it does not provide a counter-example to Bell's theorem because it does not satisfy the assumptions of that theorem.

\section{Christian's argument contra Bell}

The physical assumptions of local realism lead to a mathematical model for a Bell-CHSH type experiment where Alice and Bob each toss a coin to select a measurement setting on a measurement apparatus, and then go on to observe a binary outcome, of the following form. The outcomes of the two coin tosses will be denoted by $A$ and $B$, they take the values in the set $\{1, 2\}$. These are just labels. The observed measurement outcomes will be denoted by $X$, $Y$, they take values in $\{-1, +1\}$. In one \emph{trial} one observes one quadruple $(A, B, X, Y)$. According to local realism, these observed random variables have the following hidden structure. There exists a set of four counterfactual outcomes $(X_1, X_2, Y_1, Y_2)$ such that $X= X_A$ and $Y= Y_B$; thus the coin tosses merely \emph{select} which of the counterfactual outcomes $X_i$ and $Y_j$ actually get observed. We furthermore assume that $(A, B)$ is statistically independent of $(X_1, X_2, Y_1, Y_2)$.

We already took account of the assumption of \emph{locality} by giving $X_i$ and $Y_j$ each just one index. The outcome which Alice would have seen had she chosen setting $i$ and Bob chosen setting $j$, which one could denote by $X_{ij}$, is such that $X_{i1} = X_{i2} =: X_i$.  It is made plausible in experiments by the spatio-temporal arrangement of insertion of settings or inputs (one in each wing of the experiments) and recording of outcomes or outputs (also one in each wing of the experiment). Bob's outcome must have been observed before Alice's setting choice could have become known at Bob's location, even if travelling at the speed of light, and vice-versa.

The assumption $X = X_A$ says that the actually observed outcome on Alice's side only depends on Alice's setting $A$ and not on Bob's setting $B$. In standard probabilist's shorthand, it actually stands for $X(\omega) = X_{A(\omega)}(\omega)$ for all {$\omega\in\Omega$}, an underlying probability space on which all these random variables are defined.

The assumption of \emph{realism} is the assumption that the quadruple $(X_1, X_2, Y_1, Y_2)$ can be (mathematically) defined at all. Sometimes this assumption is called \emph{counterfactual definiteness}. A local-realistic mathematical model of the physics going on in this experiment allows one to include in the model, in a consistent way, what the outcome would have been, had certain physical settings been different from what they were in actuality.

The assumption of statistical independence between settings and counterfactuals is an assumption of \emph{freedom} or of \emph{no-conspiracy}. 

Note that in this paper, the three assumptions (locality, realism, no-conspiracy) are simple mathematical assumptions concerning the mathematical existence of a model with certain mathematical properties and which reproduces exactly certain predictions of quantum mechanics. The words have, of course, a long history and are associated with philosophical positions and past controversies concerning the philosophy of science. My terminology is nowadays pretty standard. Moreover, {``counterfactual definiteness''} has become a standard term in the modern scientific and statistical understanding of \emph{causality}. Counterfactual reasoning is common currency in medical statistics, in epidemiology. It is essential in moral and in legal reasoning, and in the understanding of (and learning from) history. Statisticians say that they can only establish correlations, not causation; but they are hired to establish causation.

A further side remark is that Bell, following EPR (Einstein, Podolsky and Rosen), originally derived realism from locality, by noting that with equal settings, outcomes were equal and opposite. It is hard to conceive that this could be the case if the outcome on each side of the experiment was not actually predetermined in some way or other. This leads furthermore to the idea that all randomness in the measurement outcomes is purely due {to} randomness at the source. However, for very sound reasons (in experiments one does not observe \emph{perfect} anti-correlation at equal settings; at best, only approximate anti-correlation) Bell and the whole community rapidly adopted the CHSH inequality and allowed for further randomness at the measurement locations.

Given these assumptions, let us take a look at the following expression $Z:= X_1 Y_1 - X_1 Y_2 - X_2 Y_1 - X_2 Y_2$. One can rewrite it as $X_1(Y_1 - Y_2) - X_2(Y_1 + Y_2)$. The four random variables $X_i$ and $Y_j$ take the values $\pm 1$, so either $Y_1 = Y_2$ or $Y_1 = - Y_2$, so one of the two terms in brackets equals zero, the other equals $\pm 2$. They are each multiplied by $\pm 1$ so the value of the whole expression is $\pm 2$. That implies that its expectation value lies between $-2$ and $+2$; in particular, it cannot exceed $+2$. By linearity of the expectation operation, $\mathbb E(X_1 Y_1) - \mathbb E(X_1 Y_2) - \mathbb E(X_2 Y_1) - \mathbb E(X_2 Y_2) \le 2$. By independence,  $\mathbb E(X_i Y_j) =  \mathbb E(X_A Y_B \mid A = i, B = j) = \mathbb E(X Y \mid A = i, B = j) =: \mathbb E_{ij}(XY)$. This leads to the following constraint on four experimentally accessible quantities, a Bell-CHSH inequality:
$$ S := \mathbb E_{11}(X Y) - \mathbb E_{12}(X Y) - \mathbb E_{21}(X Y) - \mathbb E_{22}(X Y) \le 2.\eqno (1)$$

Christian writes \emph{As innocuous as the step} [taking the expectation of $Z$] \emph{in the proof may seem mathematically, it is, in fact, an illegitimate step physically, because what is being averaged on its right-hand are unobservable and unphysical quantities. Indeed, the pairs of measurement directions $(a, b)$, $(a, b')$, $(a', b)$, and $(a', b')$ are mutually exclusive measurement directions, corresponding to incompatible experiments which cannot be performed simultaneously}. 

His remark does explain why quantum mechanics need not admit this same bound. But once one has assumed locality, realism, and no-conspiracy, the step of taking the expectation value of a quantity which does exist in the mathematical model which one is using to describe this experiment is legitimate. The Bell-CHSH one-sided four-correlation inequality (1) holds under the assumptions of locality, realism and freedom, which are inspired by classical physical thinking, and are even deeply engrained in our physical intuitions. If it is violated experimentally, then we know that experimental reality cannot be modelled in a way which is consistent with those three assumptions.

\hl{In fact, Christian's misunderstanding, that the derivation of the Bell-CHSH inequalities depends in any way on assuming that it is physically possible to measure spin in different directions simultaneously, or physically meaningful to consider what would happen if one did so, is as old as the hills. Already in 1975 Bell published a paper ``Locality in quantum mechanics: reply to critics'', answering five published critiques of his work by eminent physicists, some of whom raised exactly the same issue. The paper is included in his book of collected articles on this topic. He writes
}

\begin{quote}
\hl{These authors say `clearly since $A$, $A'$, $B$, $B'$ are all evaluated at the same $\lambda$, they must refer to four measurements carried out on the same electron-positron pair'.  [...] But we are not at all concerned with sequences of measurements on a given pair of particles,  or of pairs of measurements on a given pair of particles. We are concerned with experiments in which for each pair the spin of each particle is measured once only. The quantities [...] are just the same functions with different arguments.''.}
\end{quote}

\hl{Bell understands that in mathematical physics there is a phase at which one postulates a mathematical model for certain physical phenomena. Then, within that mathematical model, one deduces mathematical consequences. The aim is naturally to deduce consequences which can be translated back into physical predictions. That is what he did. It turned out that experiments could be done which violated consequences of the initial assumptions.}

\section{Christian's first model: Pearle's detection loophole model}

Suppose now the measurement outcomes take values in $\{-1, 0, +1\}$ where the outcome ``zero'' stands for no detection. Inequality (1) is easily shown to remain true. But an experimenter might be tempted not to estimate the correlations  $\mathbb E_{ij}(XY) :=  \mathbb E(X Y \mid A = i, B = j)$, but instead the conditional correlations $\mathbb E_{ij}(XY):= \mathbb E(X Y \mid A = i, B = j, X\ne 0, Y\ne 0)$. Pearle \cite{pearle} showed that under local realism, it is now possible for the Bell inequality (1) to be violated, if the outcome $0$ is frequent enough. Pearle's paper contains a number of misprints and mistakes, and his final model description requires some further mathematical development before it is ready for implementation as computer code, see \cite{gill2020b}.

The fact that the Pearle model existed and did exactly what it was intended to do, namely to exactly reproduce the singlet correlations, was well known. Christian states ``\emph{the 3-sphere model has nothing whatsoever to do with data rejection or detection-loophole}''.  This is easy to say, but does no justice to the facts. His computer simulation certainly does use data rejection, as anyone can see who studies Christian's code carefully. The simulation does not constitute an empirical counterexample to Bell's theorem. Due to the post-selection of data, the conditional distribution of the settings given the hidden variables in the model will depend on the hidden variables, and vice versa. Since the model is local and realistic, and since it violates Bell inequalities, it must by Bell's theorem violate \emph{no-conspiracy} -- {the statistical independence between settings and counterfactuals}.

{The easiest way to see this without studying the code in detail is by adding an extra line of code at the end of the program cited by Christian in his paper}, \url{https://rpubs.com/jjc/13965}, {in which the actual numbers of observations used to calculate each empirical correlation are exhibited. It will then be noticed that a large proportion of simulated particle pairs have gone missing. Notice that Christian starts the simulation by generating what he calls a ``pre-ensemble'' of size \texttt{M}. He later selects a subensemble called \texttt{good}. His vector \texttt{Ns} contains the number of ``good'' observations for each angle difference. One just needs to print \texttt{sum(Ns)} in order to see how many particle pairs went missing. But anyway, the line defining the subset of particle pairs \texttt{good} is clear enough evidence.}

\section{Christian's second model: Bertlmann's socks}

In Section VII, Christian uses Geometric Algebra to ``rederive'' his model. Here I can be brief, since this model is discussed by him in his companion paper \cite{IEEEAccess2} and analysed by myself in \cite{gill2021}.  His equations (66) and (67) tell us quite unambiguously that in his local hidden variables model, whatever settings $a$ and $b$ are used by Alice and Bob, the observed outcomes $X$, $Y$ will satisfy $X = -Y = \pm1=\lambda$, a fair coin toss. Whatever settings are used, the correlation predicted is $-1$, and this does not violate any of the eight one-sided four-correlation Bell inequalities. Of course not: Bell's theorem is true, and the model is local, realist, and satisfies no-conspiracy.

\section{Conclusion}

Unfortunately, Christian's paper \cite{IEEEAccess1} makes no contribution to the ongoing debates concerning Bell's theorem. It is marred by elementary errors in calculus and algebra, and exhibits fundamental errors of reasoning concerning Bell's theorem. The computer simulation exhibited in the paper is based on a well-known 50 year old detection-loophole model. Bell's inequality follows from a set of physical assumptions and if one relaxes those assumptions, then Bell's conclusion need not hold. The geometry of space plays no role.

\section*{Appendix: Objections And answers}
This appendix addresses some possible objections or concerns on various points of this paper.

Objection A: Christian's model is based on a different algebra than that he used in his paper in \emph{Royal Society Open Science}, so Lasenby's critique of that paper is irrelevant.

Answer: \hl{Christian's explicit definition of measurement functions results in measurement outcomes which are equal and opposite with probability one, \emph{whatever the measurement settings}. In all his works on Bell's theorem he makes the same elementary errors in GA. It is relevant, albeit indirectly, that an eminent specialist in GA has confirmed this.}

Objection B: Christian's contribution is to show that Bell assumed a ``flat'' space. If space is curved then Bell's approach breaks down.

Answer: Bell makes no assumptions whatsoever about the geometry of space because geometric considerations are irrelevant to his argument. In rigorous Bell-type experiments, at two distant locations, two experimenters each use a randomiser to set a switch on an apparatus. A little later, each apparatus displays ``up'' or ``down on a LED screen. The timing of these events is carefully coordinated so that Alice's outcome could not be influenced by Bob's setting through any means propagating at or below the speed of light, or vice versa.

Objection C: Physically we know that only relative orientations are physically meaningful. The orientations of measurement devices in each lab, separately, cannot be part of the fundamental physics of EPR-B experiments. 

Answer: \hl{The assumptions of local realism tested in a Bell type experiment do not exclude that only relative orientations are physically meaningful. In fact, the experiments typically confirm that the correlations between outcomes at the two measurement stations only depend on the difference between the two externally chosen measurement directions.The experiments test whether or not the observed correlations could be explained using an underlying physical description of a classical nature. The conclusion of rigorous experiments is that this is not the case.}

Objection D: Christian's work does not challenge established mathematical results in algebra or probability theory. Rather, he aims to show that conventional physical concepts of locality and reality need to be reviewed.

Answer: Of course, conventional physical concepts of locality and reality should not be taken for granted. Indeed, many approaches to the interpretation of quantum mechanics and to the resolution of paradoxes in quantum theory (a paradox is an apparent contradiction, not an actual contradiction), try to do just that.There is nothing wrong with that aim, if it is indeed Christian's main aim.  However, as a mathematician in science, it is my task to pay special attention to the mathematics he actually writes down. His verbal interpretations are interesting and sometimes helpful. But the fact remains that he has repeatedly claimed to have an original and purely mathematical counterexample to the commonly agreed mathematical core of Bell's arguments. \hl{His mathematical model has varied over the years. As a mathematician, I can confidently state that his past attempts were all failures. The present paper makes the same general claim as in his other works, but with yet another mathematical construction, and it also fails.}

Objection E: Since the work of Gödel we can no longer trust conventional mathematics. Perhaps Bell's theorem is both true and not true under the conventional or standard ZFC axioms: in other words, those axioms are inconsistent. Already, Einstein had worries that the notion of ``real number'' is probably unphysical.

Answer: Mathematicians have worked hard to show that the mathematics commonly used by physicists does not require the axiom of choice.  Researchers in the foundations of mathematics have suggested replacing the axiom of choice by an axiom inspired by physical intuition. But my own refinements of Bell's theorem (taking account of time and of finite statistics) are essentially based on discrete mathematics: counting arguments. Those who think that Bell's theorem signals a breakdown of mathematics as we know it must follow their instincts. Tim Palmer argues that p-adic topology and super-determinism are the way forward. Itamar Pitowsky argued for non-measurability. It would not surprise me if at some point it turned out that the marriage of quantum theory and relativity theory might depend on introducing radical new frameworks into mathematics. But in any case, these arguments do not rescue Christian's paper.

Objection F: A computer simulation proves nothing anyway, so why pay so much attention to Christian's Monte Carlo simulation.

Answer: Christian claims that his computer simulation is a faithful representation of his mathematical model, and claims that it does constitute proof that Bell's theorem is wrong. A faithful computer simulation of a loophole-free Bell-type experiment would in fact constitute extremely powerful direct experimental proof that classical physical systems can violate Bell inequalities even under the most stringent spatial and temporal constraints nowadays routinely enforced in the quantum optics laboratory. On a single computer one can simulate the behaviour of a network of ordinary digital computers connected by ordinary one-way digital communication lines. Monte-Carlo simulations do not generally provide mathematical proof, but they can provide overwhelming scientific evidence. Such experiments can be reproducible in the strongest sense possible, as long as they only need commonly available computational resources. One needs a PC and an internet connection and some programming experience. So Christian is to be applauded that he did include discussion of Monte-Carlo simulation of his model in his paper and made the code freely available. These are good research practices. They are meant to allow other researchers to be able to check published results. I did carefully check, and I report my findings here. Anyone can check them. 

Objection G: The fact that some formulas in Pearle's detection-loophole model coincide with formulas in Christian's model is some kind of coincidence. The particle pairs which are rejected in Pearle's model because either particle did not trigger a detector were never actually created and emitted from the source in Christian's model.

Answer: according to that interpretation, the hidden variable in Christian's model has a probability distribution which is drawn from the conditional distribution of Pearle's hidden variable, conditional on both particles being detected and given the detector settings in force. This conditional probability distribution actually depends on both detector settings. The computer simulation does embody locality and realism, but it violates ``no conspiracy''. In order to implement it directly on a computer network it would be necessary for the settings first to be communicated to the source; then the hidden variable could be generated according to the distribution just described; only then could the hidden variable travel with the two particles to the two detectors. By the way, the hidden variable in the Pearle-Christian model is not a fair coin toss used to randomly flip the orientation of space! It's a random direction in three-dimensional space, and a random length, defining a random point in the unit ball in three-space. 

Objection H: Realism is ``the presupposition of every kind of physical thinking'' rather than a claim which can be disproved with any experimental results. Bell’s theorem reveals the absurdity of quantum mechanics and the inability of most modern scientists to think logically rather than refutes realism.

Answer: That is an interesting point of view. Indeed for many scientists, local realism is an axiomatic and essential part of logical thought, it is not something which can be disputed at all. Hence if quantum mechanics violates it, then quantum mechanics is wrong, and if not, then the empirical predictions of quantum mechanics do actually have a ``mechanistic'', classical-like explanation. Some ingredients of quantum mechanics might be correct, but entanglement is not something spooky and mysterious. This conviction has inspired some of the most challenging criticism of the theory and hence has also inspired theorists to conceive of, and experimentalists to perform, ever more outrageous experiments. The end of that process is nowhere in sight. As Bohr said: now we have a contradiction, now we can at last make some progress!

\begin{IEEEbiography}[{\includegraphics[width=1in,height=1.25in,clip,keepaspectratio]{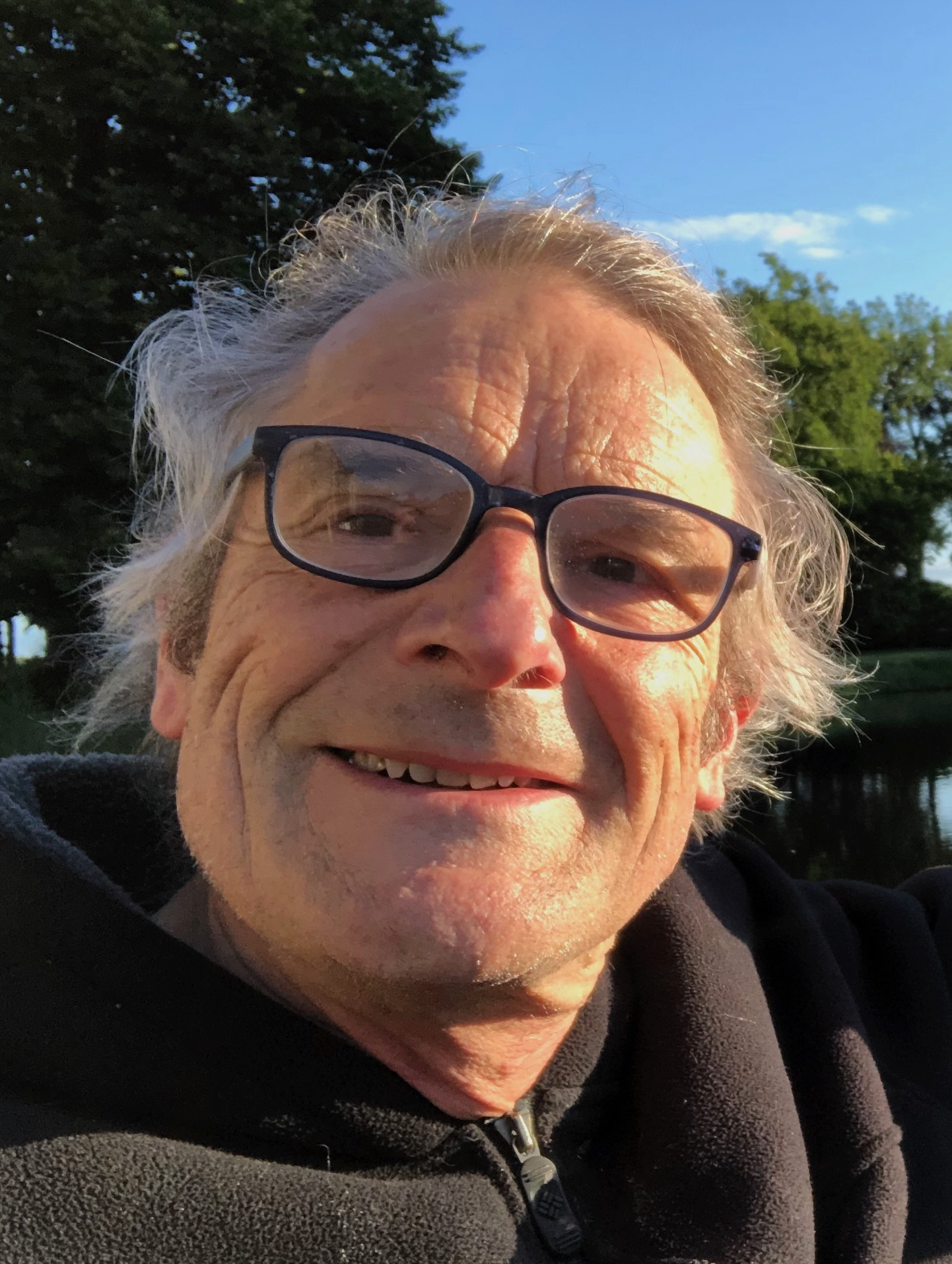}}]{Richard D. Gill} 
was born in 1951 in the UK. B.A. degree in mathematics, Cambridge University, 1973; diploma of statistics, Cambridge University, 1974; PhD degree in mathematics, Free University Amsterdam, 1979. In his career he has been head of statistics department, CWI Amsterdam; professor mathematical statistics in Utrecht, later in Leiden; and is now emeritus professor in Leiden. His early work was in counting processes, survival analysis, martingale methods, semiparametric models. Later he has worked in forensic statistics, quantum information, and on scientific integrity. His work on experimental loopholes in Bell-type experiments was incorporated in the famous ``loophole-free'' Bell experiments of 2015. He is a member of the Royal Dutch Academy of Science and a past president of the Netherlands Society for Statistics and Operations Research.
\end{IEEEbiography}

\EOD

\end{document}